\documentclass[prb,aps,twocolumn,showpacs]{revtex4}
\usepackage{epsfig}
\usepackage{amsmath}
\usepackage{verbatim}

\begin{document}

\title{High resolution angle resolved photoemission studies on
quasi-particle dynamics in graphite}

\author{C. S. Leem, Chul Kim, S. R. Park, Min-Kook Kim, Hyoung Joon Choi, and C. Kim$^*$}

\affiliation{Institute of Physics and Applied Physics, Yonsei
University, Seoul, Korea}

\author{B. J. Kim}
\affiliation{School of Physics and Center for Strongly
Correlated Materials Research, Seoul National University, Seoul,
Korea}

\author{S. Johnston$^{1,2}$, T. Devereaux$^1$}
\affiliation{$^1$Department of Photon Science, Stanford Linear Accelerator Center, Stanford University, Menlo Park, CA, 94025, USA}
\affiliation{$^2$Department of Physics and Astronomy, University of Waterloo, Waterloo, ON, Canada, N2L 3G1}

\author{T. Ohta, A. Bostwick, E. Rotenberg}
\affiliation{Advanced Light Source, Lawrence Berkeley National
Laboratory, Berkeley, California 94720, USA}

\date{\today}

\begin{abstract}
We obtained the spectral function of the graphite $H$ point using high
resolution angle resolved photoelectron spectroscopy (ARPES). The
extracted width of the spectral function (inverse of the
photo-hole lifetime) near the $H$ point is approximately
proportional to the energy as expected from the linearly
increasing density of states (DOS) near the Fermi energy. This is
well accounted by our electron-phonon coupling theory considering
the peculiar electronic DOS near the Fermi level. And we also
investigated the temperature dependence of the peak widths both
experimentally and theoretically. The upper bound for the
electron-phonon coupling parameter is ~0.23, nearly the same
value as previously reported at the $K$ point. Our
analysis of temperature dependent ARPES data at $K$ shows that the
energy of phonon mode of graphite has much higher energy scale
than 125K which is dominant in electron-phonon coupling.
\end{abstract}
\maketitle

\section{Introduction}

Fermi liquid theory\cite{Landau} (FLT) is thought to be one of the
most successful theories for describing the behaviors of electrons
in solids, especially electrons near the Fermi energy in metals at
low temperature. The success of the FLT in metallic systems
naturally raises an issue on how far the FLT scheme can be applied
to other condensed matter systems. Related to this question, there
is a long-standing controversy on whether electrons in graphite, a
2 dimensional semi-metallic system, can be described within the
FLT scheme or not. According to FLT, the lifetime of an electron due to electron-electron interactions
is inversely proportional to the square of the binding energy.
Therefore, measurement of the lifetime as a function of the
binding energy of an electron would be a direct test of the
validity of FTL in graphite.

Experimental results do not seem to show evidence for Fermi liquid
behavior of electrons in graphite\cite{Xu}. In fact, the inverse
lifetime measured by 2 photon photoemission experiments (2PPE)
conducted on natural single crystal graphite and highly oriented
pyrolytic graphite (HOPG) appears to increase linearly as a
function of the binding energy\cite{Moos}. The observed peculiar
behavior in the energy dependence of the inverse lifetime was
discussed in terms of the peculiar dispersion of plasmon\cite{Xu}
or electron-electron interaction in combination with the band
structure of graphite\cite{Gonzalez, Spataru}.

However, electron-phonon coupling (EPC), one of the most
fundamental interactions in solids, has not been considered in the
discussion. On the theoretical side, very little work can be found
on the EPC in semi-metals even though it has been well developed
and widely studied in metallic systems\cite{Grimvall}. Only very
recently has some theoretical models for graphene
appeared\cite{Park,Calandra,Lee}. On the experimental side,
electron lifetime was measured only for the energies larger than
the maximum phonon energy of graphite ($\sim$ 200
meV)\cite{Tuinstra,Maultzsch,Mohr} in the 2PPE
experiments\cite{Xu,Moos}. Therefore, to address the lifetime
issue due to EPC in graphite, one may need two requirements. First,
 the experimental data must show the lifetime of
quasi-particles sufficiently close to the Fermi energy, less than the
maximum phonon energy of graphite. Second, a proper
model that considers the electron-phonon interaction contribution
to the quasi-particle decay should be developed. In regards to the
second point, models developed for metals have been used in the
analysis of angle resolved photoemission (ARPES) data on graphite
due to the lack of theoretical EPC models for
semi-metals\cite{Sugawara2007}.

To address the issue of the quasi-particle dynamics and EPC in
graphite, we performed high resolution ARPES experiments on high
quality natural single crystal and developed a theoretical model that
considers linear density of states (DOS) near the Fermi
energy\cite{Leem}. Our previous work was performed near the
$K$-point and showed a relatively small EPC constant $\lambda$ =
0.20. To extend our previous we have
obtained high resolution ARPES data from the $H$ point to determine if it also has a
small EPC constant.
In addition, we also performed temperature dependent studies near
the $K$ point. The temperature dependent data is compared with a
theoretical model that fully considers the graphite DOS. Properly
extracted peak widths are well understood within our EPC model
with a linear DOS near the Fermi energy and shows a small EPC
constant of less than 0.23.

\section{Theory}

We first consider the theoretical side of the quasi-particle dynamics
in graphite. In this section, we discuss possible decay channels
for quasi-particles in graphite. First, it will be discussed how
the lifetime of quasi-particles in graphite can be affected by
EPC. We will formulate the self energy of quasi-particles based on
the linear DOS of graphite, for zero temperature in section 1 and
for a finite temperature in section 2. The latter is to establish
the foundation for estimating the EPC constant through temperature
dependent studies. Second, we will discuss other decay channels
such as electron-electron scattering, electron-plasmon scattering,
and impurity and defect scattering. Through these discussions, we
wish to establish that the dominant scattering mechanism for
quasi-particles in graphite comes from the EPC.

\subsection{Electron-phonon coupling in graphite}

Electron-phonon interaction theory is an extensively
studied subject in condensed matter physics. The importance of
its role is high-lighted in the theory for conventional
superconductors, i.e., the BCS theory. Even though a general
theory should be applicable to any system, specific and more
applicable models have been developed for metallic systems.
However, a key assumption used for metallic systems, constant
DOS near the Fermi level, is not valid for semi-metals and
insulators. To the best of our knowledge, EPCs in semi-metals and
insulators have not been thoroughly studied theoretically
(probably due to lack of interest). With recent developments in
graphene/graphite related research\cite{Park,Calandra,Bostwick},
EPC in semi-metals has become more important. Therefore, we need a model to
evaluate the EPC constant in graphite.

\begin{figure}
\centering \epsfxsize=8.5cm \epsfbox{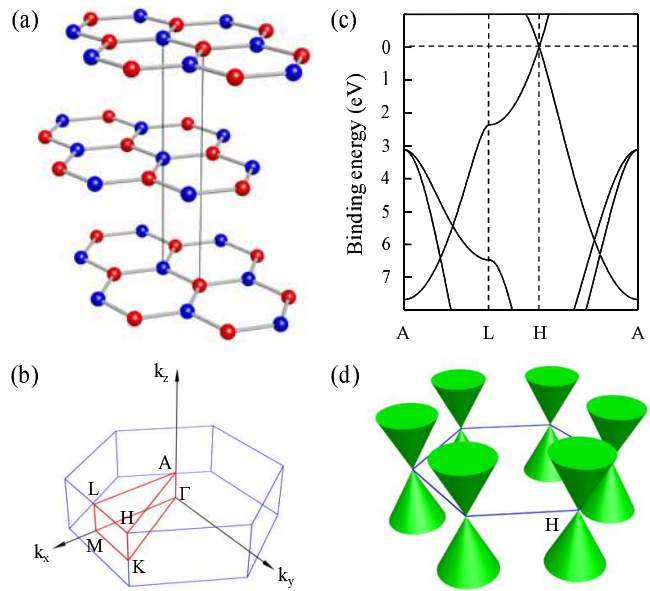} \caption{(a)
The structure of graphite. Spheres are carbon atoms. Graphite
shows a layered structure, which has an stacking order of
\textit{ABAB}... (b) First Brillouin zone of graphite. The
symbols represent high symmetry points. (c) Calculated electronic
band structure along $A$-$L$-$H$-$A$. (d) Approximated band
structure. Each corner of hexagon is $H$ point and $z$ direction
is energy. } \label{fig1}
\end{figure}

To understand the EPC in graphite, one should consider it's
characteristic band structure near the Fermi level.
Fig. 1(a) shows the crystal structure of graphite. Graphite has a
layered structure and the stacking order is \textit{ABAB...} In
each layer, carbon atoms form strong $\sigma$ bonds produced by
sp$^2$ hybridization while the out-of-plane $p_z$ orbitals form
$\pi$ bonds. Fig. 1(b) depicts the first Brillouin zone (BZ) and
high symmetry points of graphite in reciprocal space. The
calculated electronic band dispersion of graphite along the high
symmetry line, A-L-H-A, is plotted in Fig. 1(c). The band
dispersion within $\pm$1 eV near the H point is almost linear. A
three dimensional view of the band dispersion is shown in Fig.
1(d). The point at which the two cones meet each other is at the Fermi
energy and is called the Dirac point. This band structure yields a DOS
which increases linearly with binding energy (linear DOS).
Numerous studies of the band structure of graphite can be found
both theoretical\cite{McClure, Slonczewski, Zunger, Tatar,
Charlier} and experimental sides\cite{Law1983, Takahashi, Law1986,
Collins, Maeda, Vos, Shirley, Heske, Strocov, Kihlgren, Zhou2005,
Sugawara2006, Zhou2006-1, Zhou2006-2}.

\begin{figure}
\centering \epsfxsize=8.5cm \epsfbox{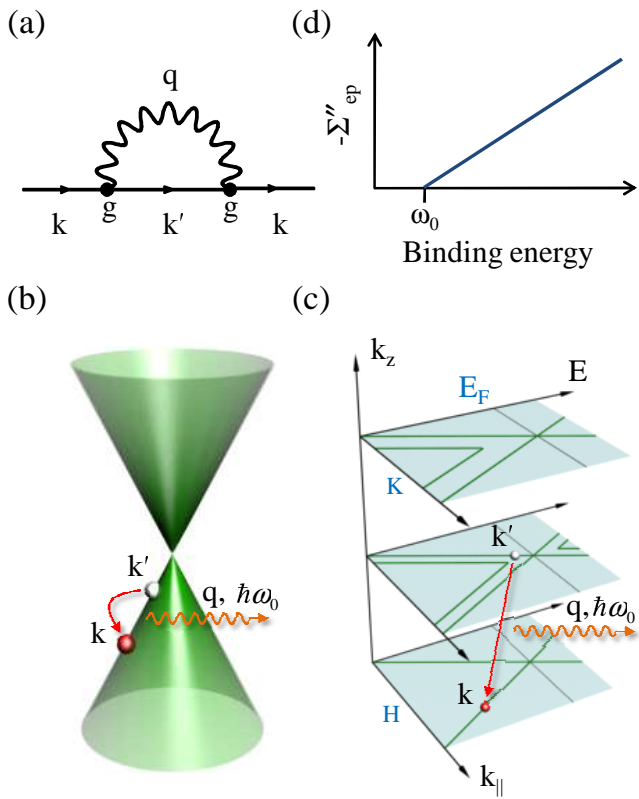} \caption{(a)
The lowest order Fyenman diagram for EPC. $g$ is electron phonon
coupling constant. $k$ and $k^\prime$ are crystal momenta of
holes. $q$ is momentum of phonon. (b) Schematic diagram of the EPC
process as shown in panel (a). Photo-hole $k$ makes a transition
to $k^\prime$ emitting phonon of $q$. $\hbar\omega_0$ is the
emitted phonon energy. (c) Schematic diagram for scattering in
$k_z$ direction. (d) The imaginary part of the self energy vs. binding
energy predicted by our qualitative theory (see the
text).} \label{fig2}
\end{figure}

The electron-phonon coupling theory in graphite should be
considered within this characteristic linear DOS. To understand
the electron-phonon coupling in graphite, one needs to get the
real or imaginary parts of the self energy. If one approaches
electron-phonon coupling through the real part of the self energy,
one has to confront a task of finding the bare band. The bare band
of graphite is not linear and hence much harder to guess in
comparison with metallic systems. Even though it was argued that
the experimentally measured band structure at $H$ is
linear\cite{Zhou2006-1}, our results show that the dispersion is
not linear and has some parabolic character near the Fermi energy.
Therefore, we chose to use the imaginary part in the analysis.
Note that the real part of the self energy can be obtained by
Hilbert transforming the imaginary part. Fig. 2(a) shows the
Feynman diagram\cite{Binosi} for the lowest order EPC under
consideration. Our model considers only this lowest order EPC in this
section. One can describe the many-body effects on quasi-particles
using a self energy scheme. The imaginary part of self energy is
proportional to the scattering rate of quasi-particle. Here, we
present the EPC process for zero ($T=0$) and finite temperature
cases ($T\neq0$) separately.

\subsubsection{Zero temperature case}

The Hamiltonian of EPC interaction can be
written as,
\begin{equation}
H_{\text{ep}}=\sum_{i,\sigma,\nu}g_{\nu}c^{\dagger}_{k+q,\sigma}c_{k,\sigma}(b^{\dagger}_{q,\nu}+b_{-q,\nu})
\end{equation}
where $c^{\dagger}_{k,\sigma}$ ($c_{k,\sigma}$) creates
(annihilates) an electron with spin $\sigma$ and momentum $k$
while $b^{\dagger}_q$ ($b_q$) creates (annihilates) a phonon $\nu$
with momentum $q$. The scattering amplitude $g$ is taken to be
energy and momentum independent. $b^{\dagger}_{q,\nu}$ term is for phonon emission process and $b_{-q,\nu}$ term is for phonon absorption. Then the imaginary part of the self-energy is
defined as a convolution over the density of states:\cite{Mahan}
\begin{equation}
\begin{split}
\Sigma^{\prime\prime}_{\text{ep}}(\omega)=\sum_{\nu}-g^2_{\nu}\pi
  (&\mathcal{D}(\omega-\omega_\nu)[f(\omega_{\nu}-\omega)+b(\omega_{\nu})]
\\+&\mathcal{D}(\omega+\omega_{\nu})[f(\omega_{\nu}+\omega)+b(\omega_{\nu})])
\end{split}
\end{equation}
where $\mathcal{D}$ is the electronic DOS and $\omega_{\nu}$ is
the energy of a phonon $\nu$. $f$ and $b$ are Fermi-Dirac and
Bose-Einstein distributions, respectively. Since the electron-phonon interaction does not alter the spin of a conduction electron, spin index $\sigma$ is suppressed to consider only one spin direction.

If we assume an Einstein phonon with an energy of $\omega_0$ and
momentum independent coupling amplitude $g$, Eqn. (2) becomes
\begin{equation}
\begin{split}
\Sigma^{\prime\prime}_{\text{ep}}(\omega)=-g^2\pi(&\mathcal{D}(\omega-\omega_0)[f(\omega_0-\omega)+b(\omega_0)]
\\+&\mathcal{D}(\omega+\omega_0)[f(\omega_0+\omega)+b(\omega_0)])
\end{split}
\end{equation}

For the zero temperature case, Fermi-Dirac function can be replaced by step function and the Bose factor is zero in Eqn. (3). Then, Eqn. (3) can be written as,
\begin{equation}
\begin{split}
\Sigma^{\prime\prime}_{\text{ep}}(\omega,T=0)=-g^2\pi[&\mathcal{D}(\omega-\omega_0)\Theta(\omega-\omega_0)
\\+&\mathcal{D}(\omega+\omega_0)\Theta(-\omega-\omega_0)]
\end{split}
\end{equation}
where $\Theta$ is a step function, $\Theta(x)=0(x<0)$ and $\Theta(x)=1(x\geq0)$.

We assume a conical band structure with the Fermi energy at the
apex of the cone. There is another conical band above the Fermi
energy which is unoccupied, and these two conical bands form a
Dirac-cone-like band structure as shown in Fig. 2(b). If a
photo-hole with momentum $\mathbf{k}$ (filled circle) is created
by a photon as shown in Fig. 2(b), it can be filled by an electron
with energy of $\omega_{k^\prime}=\omega_k-\omega_0$ and momentum
$\mathbf{k^\prime}=\mathbf{k}-\mathbf{q}$ (empty circle) where
$\mathbf{q}$ is the phonon momentum. The scattering rate is
proportional to the number of such $\mathbf{k^\prime}$ states,
thus the DOS at $\omega_k-\omega_0$. Note that if the binding
energy of the photo-hole is smaller than the phonon energy
$\omega_0$, the scattering cannot occur because there are no
electrons with sufficient energy to emit a phonon with energy
$\omega_0$. Therefore, the imaginary part of self energy of
photo-hole as a function of the binding energy is proportional to
$\mathcal{D}(\omega_k-\omega_0)$ and looks like a schematic shown
in Fig. 2(d). Note that it is also possible that a phonon may
scatter a photo-hole in the $k_z$ direction as shown in Fig. 2(c).
The outcome is not much affected by the $c$-axis scattering due to
the weak dispersion of the $\pi$ band along $k_z$ direction.

Once $\Sigma^{\prime\prime}$ is obtained, one can obtain the real
part of self energy $\Sigma^\prime_{\text{ep}}$ by Hilbert
transforming $\Sigma^{\prime\prime}$. The electron-phonon coupling
parameter, $\lambda$ is defined as,
\begin{equation}
\begin{split}
\lambda=-\frac{\partial\Sigma^\prime_{\text{ep}}(\omega)}{\partial\omega}\biggr|_{\omega=0}
\end{split}
\end{equation}
At the $K$ point, the bonding and anti-bonding bands are split because
of the inter-layer interaction of graphite\cite{Feuerbacher}.
Considering the small inter-band scattering of the photo-hole by
a phonon between bonding and non-bonding bands at K point, the above
self energy can be extended to the double band case at $K$. This
double band case was investigated in our previous work\cite{Leem}.

We also note that $\Sigma^{\prime}_{\text{ep}}$ is not affected
seriously by the detailed shape of
$\Sigma^{\prime\prime}_{\text{ep}}$ near $\omega=0$ because
$\Sigma^{\prime\prime}_{\text{ep}}$ increases linearly. This
aspect was considered in calculating
$\Sigma^{\prime\prime}_{\text{ep}}$ for $K$ and $H$
points\cite{Lee}. It was argued that
$\Sigma^{\prime\prime}_{\text{ep}}$ is somewhat different at $K$
and $H$ because the band structure at $K$ is parabolic near the
Fermi level while that at $H$ is linear. Meanwhile some difference
between $K$ and $H$ certainly exists that affects the detailed
shape of $\Sigma^{\prime\prime}_{\text{ep}}$ near the Fermi
energy, the effect on the EPC constant $\lambda$ should to be
negligible because the contribution comes mostly from the high binding
energy side.

\subsubsection{Finite temperature case}

We now move onto the finite temperature case. In the case of
metals, there is an easy way to extract the EPC constant $\lambda$
from temperature dependent data through a simple
formula\cite{Grimvall,Balasubramanian}. The formula is derived
under the assumption that the electronic DOS near the Fermi energy
is constant, which is not the case for graphite. Here, we
investigate the temperature dependence of
$\Sigma^{\prime\prime}_{\text{ep}}$ theoretically to determine if one
can easily extract $\lambda$ from the temperature dependence data.
It turns out that a simple formula such as the one for metals can
not be formulated. However, we show some possibility of estimating
EPCs from the temperature dependent data.

The imaginary part of the self-energy by electron-phonon coupling at finite temperature was shown in Eqn. (3) of the previous section. Note that for the high phonon frequency (for example, $A_1^\prime$
or $E_{2g}$ mode in graphite) the Bose factors can be neglected
for the temperature range over which we performed our experiments (10K-225K). We consider not only these high energy phonons but also low energy phonons. Unfortunately, the temperature dependence of
$\Sigma^{\prime\prime}_{\text{ep}}$ in Eqn. (7) cannot be reduced
to a simple form as the one for a metal\cite{Grimvall} and
extracting the $\lambda$ from the temperature dependence of
$\Sigma^{\prime\prime}_{\text{ep}}$ is not an straight forward task. However, one can still obtain information from the temperature dependent data. If the low frequency phonon mode participates in EPC, the temperature dependence of $\Sigma^{\prime\prime}_{\text{ep}}$ near the Fermi level should be strong while the high frequency phonon modes should contribute little to the temperature dependence. In addtion, the Bose factor in Eqn. (3) is not negligible and $\Sigma^{\prime\prime}_{\text{ep}}$ will show clear difference at different temperature. Therefore, once the $g$ value is known, one can roughly identify which phonon mode contributes the most to EPC by fitting the temperature dependent data.


\subsection{Electron-electron interaction in graphite}

\begin{figure}
\centering \epsfxsize=8.5cm \epsfbox{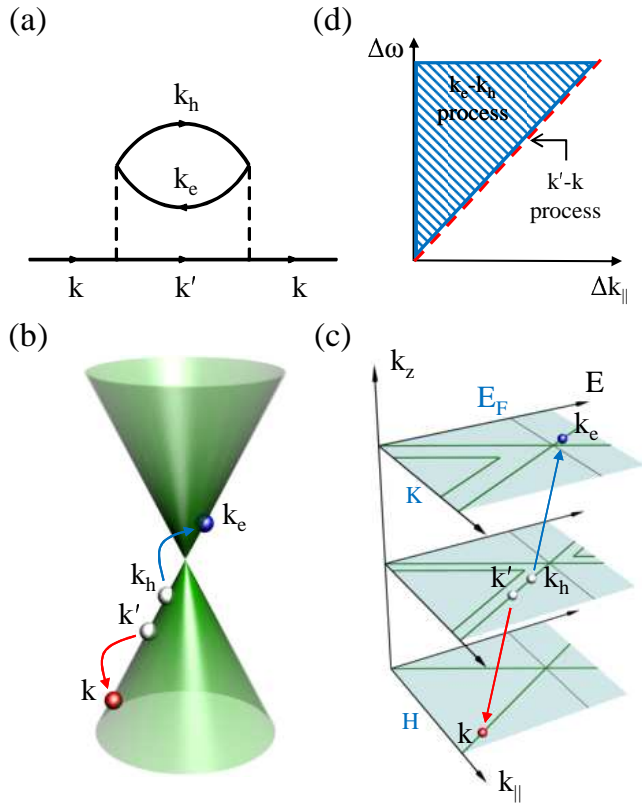} \caption{(a)
The Feynman diagram for the lowest order for electron-hole pair
creation process. Photo-hole $k$ makes a transition to a hole
$k^\prime$, and an electron $k_e$ and a hole $k_h$ are created
conserving momenta and energy. (b) Schematic diagram for
electron-hole pair creation in phase space. (c) Schematic diagram
for electron-hole pair creation considering the scattering along
$k_z$ direction. (d) Hatched area represents the possible $\Delta
k$ and $\Delta \omega$ for electron-hole pair creation. Dashed
line represents the possible $\Delta k$ and $\Delta \omega$ for
photo-hole transition. These two area slightly touch each other at
line, therefore there is no available phase space. } \label{fig3}
\end{figure}

Electron-electron interactions can also affect the quasi-particle
lifetime or the imaginary part of self energy. We consider the
lowest order scattering in electron-electron interaction occurring
via electron-hole pair creation. The Feynman diagram for this
scattering process is shown in Fig. 3(a). When the photo-hole, of
crystal momentum is $\mathbf{k}$, is created, the hole
makes the transition to $\mathbf{k^\prime}$ by creating another
hole, $\mathbf{k_h}$, and electron, $\mathbf{k_e}$. Fig. 3(b)
shows this electron-hole pair creation process in the $E-k$ phase
space. From the figure, one can see that the electron-hole pair
creation process is negligible under the linear DOS of graphite
near the Fermi energy. Let us consider the available phase space
for electron-hole pair creation in Fig. 3(b). If we plot the
energy difference ($\Delta\omega=\omega_{k^\prime}-\omega_k$ where
$\omega_{k^\prime}$ and $\omega_{k}$ are the energies of the holes
with $\mathbf{k^\prime}$ and $\mathbf{k}$, respectively) as a
function of momentum difference
($\Delta\mathbf{k}=\mathbf{k^\prime}-\mathbf{k}$), the possible
transitions occupy the area below the dashed line in Fig 3(d). In
a similar way, one can find that electron-hole creation process
occupies the hatched area in Fig. 3(d). Photo-hole decay through
the electron-hole pair creation can occur only when the two
conditions are met, that is, where the phase spaces for the two
processes overlap. They overlap only on the dashed line as can be
seen in Fig. 3(d). Therefore, the available phase space volume for
the decay through electron-hole pair creation is zero. Note that,
this is true only near the Fermi energy where the band structure
can be approximated by Dirac cones. For the photo-holes at higher
binding energies, the available phase space volume is becomes
non-zero. This fact was previously pointed out by \textit{Moos et
al.}\cite{Moos} If we limit our discussion to the low
energy dynamics in graphite, the effect of the electron-electron
interaction can be neglected.

\subsection{Other scattering mechanisms}

There are other mechanisms in graphite that may contribute to the
quasi-particle scattering such as plasmons, impurities and
defects. \textit{Xu et al.} suggested that plasmons may be the
main source for the quasi-particle scattering in
graphite\cite{Xu}. However, \textit{Spataru et al.} showed that
elctron-hole pair creation should be a more dominant mechanism than
electron-plasmon interactions for electron scattering in
graphite\cite{Spataru}. Since we have shown in our earlier
work\cite{Leem} that electron-phonon interaction is more dominant
than electron-hole pair creation based on a phase space argument,
we may conclude that electron-plasmon interaction is much weaker
than electron-phonon coupling and thus may be neglected. Impurity and
defect scattering can also contribute to the scattering rate in
graphite. These scattering mechanisms also have a rate that is
proportional to the electronic DOS as in the electron-phonon
coupling case and thus increase the slope for the imaginary part
of the self energy. This fact tells us that, if one wants to study
the electron-phonon coupling, the experiment should be conducted
on clean single crystalline graphite. In our case, we used natural
graphite single crystals which have superior quality to crystals
used in other experiments. As a results, we did not observe any
defect related states\cite{Sugawara2007,Zhou2006-2} and we
therefore believe that defect or impurity scattering is minimal.

In short electron-phonon coupling should be mechanism,
other mechanisms are suppressed due to lack of phase
space (electron-hole pair) or high quality of the crystal (low
impurity/defect levels).

\section{Experiment}

ARPES experiments were performed at Beamline 7.0.1 of the Advanced Light
Source. We used very high quality natural graphite single crystals
with sizes larger than $\approx$ 1 cm. Samples were cleaved
repeatedly \textit{ex situ} by taping method until a flat surface
without large flakes were obtained. Samples were subsequently
introduced to the ultra high vacuum chamber and annealed at
900$^{\circ}\mathrm{C}$ for 30 minutes in a vacuum better than
6.0$\times 10^{-10}$ Torr. to clean the surface. The energy
resolution was $\approx$ 40 meV. The chamber
pressure was better than 5.0$\times 10^{-11}$ Torr. during the
measurements. We found that typical size of the flat regions
without flakes was smaller than 200 $\mu$m. Therefore, we
exploited the small beam spot ($\approx$ 50 $\mu$m) to probe flat
region.

We took $k_z = H$ data at 20K with a photon energy of 103.4 eV to
obtain the electron-phonon coupling by analyzing the peak width as
a function of the binding energy. This is essentially the same as
what we reported earlier\cite{Leem} but at the $H$ point. In
addition, we performed temperature dependent experiment at the $K$
point with a photon energy of 85 eV. ARPES data was taken at 25K,
75K, 125K, 175K, 225K. We started measuring at 225K and lowered
the temperature. After having measured at 25K, we annealed the
sample again for $\approx$ 30 seconds, at $\approx$
900$^{\circ}\mathrm{C}$ and measured again. Comparison of the data
before and after annealing showed essentially no difference,
indicating there was no surface contamination during the
measurement. For comparison, graphene data were also taken at the
$K$ point. The graphene sample was epitaxially grown on 6H-SiC
{\it in situ} as reported elsewhere\cite{Ohta}. Electronic band
structure calculation was done by using the SIESTA code based on
pseudo-potential method.

\begin{figure}
\centering \epsfxsize=8.5cm \epsfbox{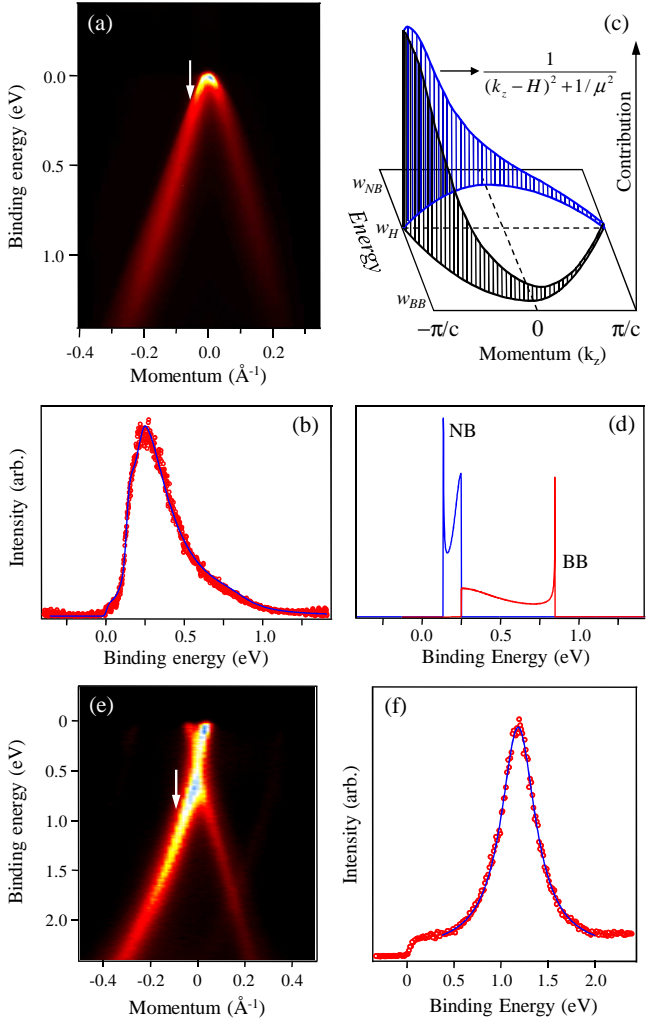} \caption{(a)
ARPES data taken at H point of graphite along $L$-$H$-$A$
direction. (b) The EDC at the $k$ point marked the arrow in panel
(a). Circles are the experimental data and thick line is the fit
for which finite escape depth effect in photoemission process has
been considered (see the text). (c) Contributions from different
$k_z$ points due to the finite escape depth effect. (d) The model
fitting function with finite escape effect considered but without
lifetime effect. (e) ARPES data at $K$ from epitaxially grown
graphene on 6H-SiC. (f) The EDC (at the $k$ point marked by the
arrow in panel (e)) shows symmetric lineshape unlike that from
graphite. The EDCs can be fitted with a single Lorentzian and
constant background (thick line).} \label{fig4}
\end{figure}

\section{Results and Discussion}

\subsection{Low temperature case}

Fig. 4(a) shows measured ARPES spectral function along the
$L$-$H$-$A$ symmetry line. The non-bonding band (NB) and bonding band (BB) are degenerate at the $H$
point whereas they are split at $K$ point. We could
identify only one peak in the energy (EDCs) and momentum (MDCs)
distribution curves. We also took data with different photon
energies, to insure that we were really at the $H$ point. The
electronic band near the Fermi energy shows a linear dispersion as
predicted in the band calculation in Fig 1.(c). However, we also
note that the band dispersion very near the Fermi level shows some
parabolic component contrary to what is expected from the theory.
This could be due to $k_z$ broadening caused by the finite escape depth. We
also note that there are no evidence defect-induced states that were
reported earlier\cite{Sugawara2006}. This indicates that our
natural graphite single crystals are of very high quality. Almost, the
negligible background of our data even at high binding energy
further supports the high quality of our sample. This means that
defect or impurity contribution to the scattering rate is very
small and we may only consider the electron-phonon coupling
effect.

Fig. 4(b) shows the EDC from the $k$ point indicated by the arrow
in panel (a). The line shape of the EDC is very asymmetric. As was
the case for the $K$ data\cite{Leem}, we can understand this
asymmetry as follows. Though we tuned the photon energy to probe
the $H$ point of graphite, the finite escape depth of the
photoelectron yields an uncertainty in $k_z$, $\Delta k_z=1/\mu$
where $\mu$ is the escape depth. Therefore, there is contribution
from other $k_z$ values which is illustrated in Fig. 4(c). As the
BB and NB have finite $k_z$ dispersions, the contribution from
other $k_z$ values results in broadening of the spectral function.
The fact that the BB has more $k_z$ dispersion gives more
broadening on the higher binding energy side as is seen in Fig.
4(b).

Fig. 4(d) depicts a model spectral function when all these effects
are accounted for. Only when such effects are considered, can one
extract the true lifetime broadening. We used $\mu=7 {\text\AA}$
for the fitting \cite{Tanuma}, and the model function in Fig. 4(d)
is convolved with a Fermi function and a Voigt function with the
Gaussian width set to the total energy resolution of 40 meV. In
addition, we introduced binding energy dependent Lorentzian width
for the Voigt function considering the observation from the $K$
data that the Lorentzian width linearly increases as a function of
binding energy\cite{Leem}.

Even though this $k_z$ uncertainty is a general property of ARPES
measurement, perfect 2D material such as graphene should not show
this escape depth effect in their ARPES data because it has no
dispersion in the $k_z$ direction. To ensure that the asymmetric
line shape in graphite is indeed from the finite escape depth
effect, we took graphene ARPES data and check if the line shape is
symmetric as expected. Fig. 4(e) shows ARPES data along the
$M$-$K$-$\Gamma$ direction of graphene which was epitaxially grown
on 6H-SiC. Fig. 4(f) is an EDC curve from the $k$ point arrow marked
in panel (e). The EDC shows very symmetric line shape contrary to
the EDC in panel (b). One can fit this curve with a single
Lorentzian with constant background as shown with the thick line
in Fig. 4(f). An almost perfect fit strongly supports the idea
that the asymmetric line shape of graphite data is indeed from the
finite escape depth effect.


\begin{figure}
\centering \epsfxsize=8.5cm \epsfbox{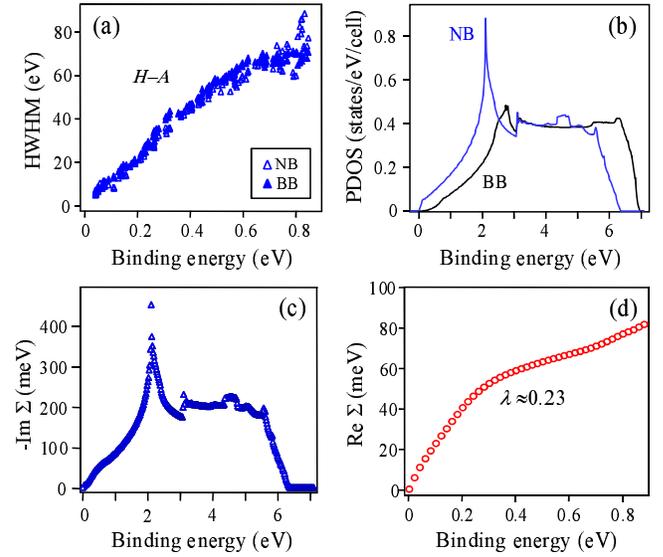} \caption{(a)
Extracted HWHM as a function of the binding energy for NB
(triangles) and BB (filled triangles). (b) Calculated pDOS for NB
and BB (c) Constructed $\Sigma^{\prime\prime}$  from the data in
panel (a) for the low energy region and pDOS in panel (b). pDOS is
scaled so that it matches the experimental HWHM at 0.9 eV. (d)
$\Sigma^{\prime}$ using Hilbert transform of
$\Sigma^{\prime\prime}$. The EPC parameter is $\approx$ 0.23.}
\label{fig5}
\end{figure}

Fig. 5(a) shows the half width at half maximum (HWHM) found by
fitting our model to the EDCs along the high symmetry line $H$-$A$.
Filled and empty symbols represent BB and NB widths, respectively.
There is almost no difference between the BB and NB widths. The
width increases linearly as a function of the binding energy. We
find that the width shows no high order dependence such as
$\sim\omega^2$. This also indicates that the EPC is the dominant
decay channel in graphite as expected from our model. Yet,
observation of very weak or no kinky feature at the optical phonon
energy of 0.2 eV shows that EPC is very weak in graphite. On the
other hand, the width converges to zero as the binding energy goes
to zero, which means that momentum mixing due to impurity or
defects is minimal, supporting again the high quality of the
samples.

One can extract the EPC parameter from the derivative of
$\Sigma^{\prime}$ at $\omega$=0. Conventionally, one obtains the
$\Sigma^{\prime}$ from the difference between the experimental
dispersion and the bare band. In graphite, this is a difficult
task because the bare band may not be linear. On the other hand,
even though harder, one can get $\Sigma^{\prime}$ by Hilbert
transforming $\Sigma^{\prime\prime}$ . To do the Hilbert
transformation, we need to know $\Sigma^{\prime\prime}$ over the
entire energy range. As this is not the case, we use scaled
partial electronic DOS (pDOS) as $\Sigma^{\prime\prime}$, assuming
that $\Sigma^{\prime\prime}$ is approximately proportional to
pDOS\cite{Leem}. Fig. 5(b) shows pDOS of NB and BB. Fig. 5(c) is
the $\Sigma^{\prime\prime}$ for NB band, obtained from the
experimental data and calculated pDOS for NB. Hilbert transform of
it gives the $\Sigma^{\prime}$ shown in Fig. 5(d). According to
Eqn. (5), we can find the electron phonon coupling parameter from
the energy derivative of $\Sigma^{\prime}$ at $\omega=0$. The
resulting value is $\lambda\approx$ 0.23, very similar to the
value of $\lambda=0.2$ for the $K$-$\Gamma$ direction reported in
our previous work\cite{Leem}. This value is larger than the
calculated value of 0.075 for graphene\cite{Park} but much smaller
than the previously reported value for
graphite\cite{Sugawara2007}. In addition, this value is consistent
with the value of 0.21 calculated with a reasonable scattering
amplitude $g$\cite{Lee}. Therefore, we conclude that EPC constant
$\lambda$ is also small at the $H$ point.

\subsection{Finite temperature case}

\begin{figure*}
\centering \epsfxsize=17.8cm \epsfbox{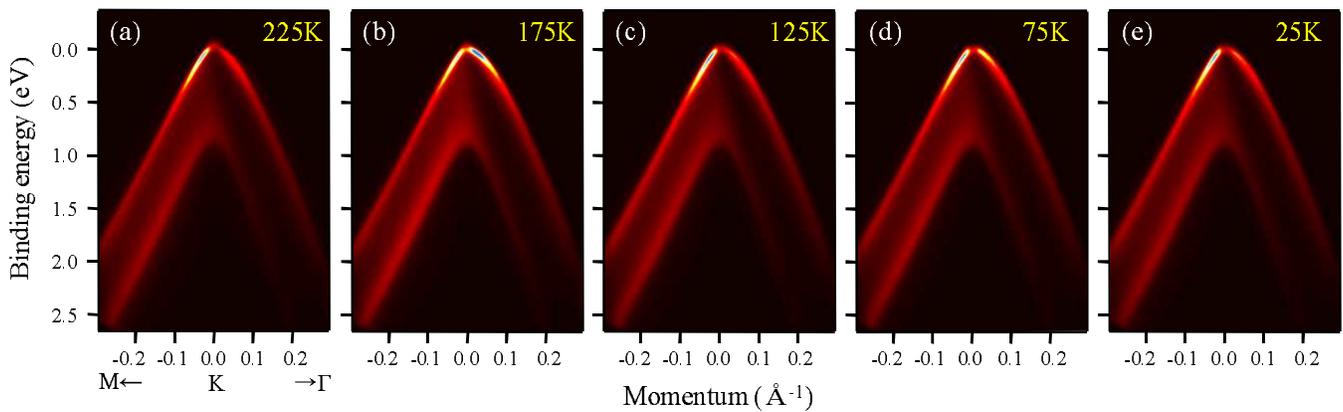}
\caption{ARPES data taken along $M-K-\Gamma$ at different
temperatures. (a),(b),(c),(d), and (e) were taken at 225K, 175K,
125K, 75K, and 25K, respectively. } \label{fig6}
\end{figure*}

Fig. 6 shows temperature dependence ARPES data at $K$ point of
graphite, which were taken at 225, 175, 125, 75, and 25K. One can
clearly distinguish the NB from the BB in each panel. Note that the
binding energy difference between the NB and the BB is about 0.8 eV. Overall, the data do not appear
to show much temperature dependence. To see this quantitatively,
we performed the same lineshape analysis we developed on the data.
Every EDC from -0.2 to 0 {\AA} of each panel in Fig. 6 is fitted
with our model function and HWHM is extracted.

Extracted HWHM vs. binding energy at different temperature is
plotted in Fig. 7. Overall, HWHMs linearly increase proportional
to binding energy. All HWHMs are quite similar to each other and
one can safely say that there is no clear temperature evolution of
spectral function. This already indicates that the energy of the
phonon mode which is involved in electron-phonon coupling in
graphite is very high compared to the temperature scale of our
measurement 225K.

\begin{figure}
\centering \epsfxsize=8.5cm \epsfbox{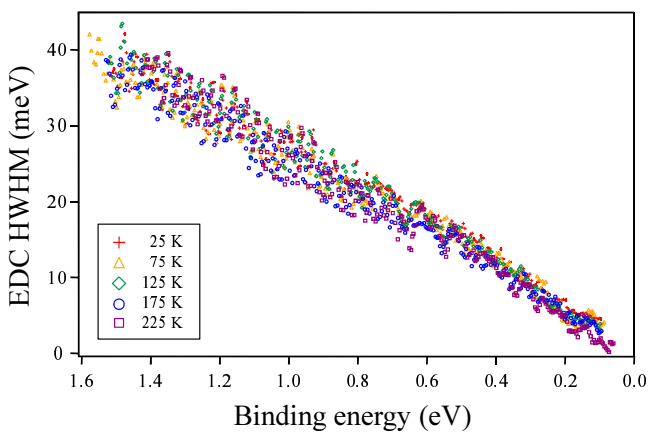} \caption{HWHM
vs. binding energy at different temperatures, 25, 75, 125, 175,
and 225K. HWHM is extracted from NB band along $M$-$K$ near Fermi
energy.} \label{fig7}
\end{figure}

Calculated imaginary part of self-energy is used to fit the
extracted HWHM of the 25K data (Fig. 8(a)). In fitting the data,
we assumed an Einstein phonon of $\omega_0$ = 200 meV. The
imaginary part is supplemented with a constant plus an
energy-dependent term $\Sigma_{\text{ee}}=A\omega^2$ in order to
simulate the electron-electron interactions. The partial DOS of the NB
band were calculated from the LDA calculation. The coupling
amplitude $g$ is a fitting parameter. We find that $g$ is
$\approx$0.39 eV and electron-electron interaction pre-factor is
$\approx$0.004. Note that negligible electron-electron interaction
near the Fermi energy is confirmed as predicted earlier in our
model.

\begin{figure}
\centering \epsfxsize=8.5cm \epsfbox{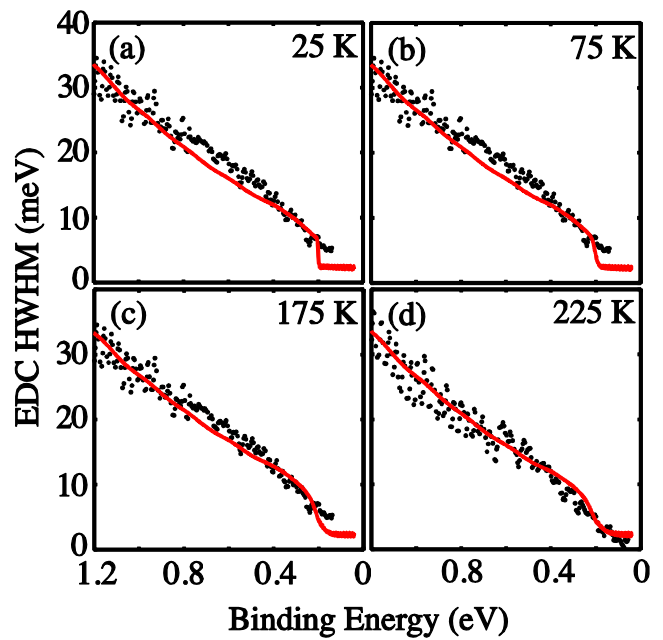} \caption{HWHMs
extracted from ARPES data at different temperatures are fitted by
calculated imaginary part of self-energy. Panel (a),(b),(c) and
(d) are for temperatures of 25, 75, 125 and 175K, respectively.
The solid line in each panel is the best fit to the experimental
data by calculated imaginary part of self-energy. The calculated
imaginary part of self-energy includes electron-phonon and
electron-electron interaction terms. The electron-phonon coupling
constant and electron-electron interaction pre-factor were used as
fitting parameters (see the text).} \label{fig8}
\end{figure}

With the fit result, one may try to evaluate the electron-phonon
coupling constant $\lambda$ from the $g$ value. By Hilbert
transforming the model fit function of $\Sigma^{\prime\prime}$, we
obtain $\Sigma^{\prime}$. Derivative of $\Sigma^{\prime}$ at
$\omega$ = 0 as in Eqn. (4) gives us $\lambda$ of $\approx$0.14.
This value is smaller than the value of $\lambda$ = 0.2 at $K$
which was obtained from the experimental data\cite{Leem}. A key
difference between the two methods is that while we assume an
Einstein phonon of $\omega$ = 200 meV, no such assumption is used
in transforming the experimental data. However, the experimental
data is more susceptible to systematic errors, especially at very
low binding energy range (where the line shape is affected by
Fermi function). Since the low energy range has more effect on
$\lambda$ and a theoretical result shows $\omega$ = 200 meV is the
dominant phonon\cite{Park}, $\lambda$ = 0.2 probably gives us the upper bound
for the electron-phonon coupling.

Other panels in Fig. 8 show HWHMs and fitted model function at
different temperatures. The fitting was conducted on NB bands
along $M$-$K$ near the Fermi energy at each temperature. As could
already been seen in Fig. 7, fitting of the HWHM results in
negligible temperature dependence. This indicates that the phonon
involved in the coupling has much higher energy scale than 225K.
In fact, the 200 meV bond-stretching mode may be the most dominant
one as the {\it ab initio} calculation on graphene
shows\cite{Park}.

\section{Conclusion}
We present high resolution ARPES data taken at the H point of
natural graphite single crystals. The graphite bands shows a linear
dispersion as predicted in LDA calculation and the NB and BB are
degenerate. First, we considered various scattering mechanisms in
graphite. We deduced theoretical formula for the scattering rate
by phonons. We find that the scattering rate by EPC increases linearly with
binding energy due to the linear density of states.
Electron-electron interactions in graphite are negligible in the low
binding energy region where the band dispersion is linear.
The impurity or defect scattering rate is also proportional to binding
energy because of linear DOS. We show that all effects
other than phonon scattering are negligible.  Second, with the
finite escape depth effect in photoemission process considered, we
extracted $\Sigma^{\prime\prime}$ from the EDCs of the NB and BB
separately. Finally, we approximated $\Sigma^{\prime\prime}$ by
combining the experimental HWHM $\Sigma^{\prime\prime}$ and
calculated partial DOS. The obtained $\Sigma^{\prime\prime}$ is
converted to $\Sigma^{\prime}$ through a Hilbert transform.
The extracted EPC parameter at $H$ is $\approx$ 0.23, which is small,
consistent with the value 0.2 from the K point in our previous work.
This small EPC parameter is also consistent with very weak kinky
features in our data.

In addition, we conducted temperature dependent ARPES measurements
on the graphite K point. The temperature dependent data shows no
notable evolution in the EDC lineshape within the temperature range
(25-225K). Analyzing the experimental temperature dependence of peak
widths and simulated temperature dependence, we conclude that the
dominant phonon mode in EPC in graphite is much larger than the
temperature scale of our experiment 225K. This is consistent with
the notion that the phonon mode in electron-phonon coupling in
graphite is the 200 meV optical phonon mode as is the case for
graphene.

Even though electron-phonon coupling has been heavily studied,
most of these studies were focused on metallic systems where the
density of states near the fermi level is approximately constant. Such is not generally
true, especially for semi-metals. The formulas discussed in this
work are very general and can be used for any shape of electronic
density of states. It should therefore be useful in the future
studies on semi-metals.

\section{Acknowledgement}

Authors acknowledge fruitful discussions with J.H.Han. This work
is supported by the KICOS in No. K20602000008. C.S.L. acknowledges support through the
BK21 Project and helpful discussions with J.-W.Rhim. H.J.C. acknowledges supports from the KRF (KRF-2007-314-C00075), the KOSEF Grant No. R01-2007-000-20922-0, and KISTI Supercomputing Center (KSC-2008-S02-0004). ALS is operated by the DOE's Office of BES.


\begin{thebibliography}{40}

\bibitem[*]{Corres} Electronic address: cykim@phya.yonsei.ac.kr

\bibitem{Landau} L. D. Landau, Sov. Phys. JETP {\bf 3}, 920 (1957).

\bibitem{Xu} S. Xu, J. Cao, C. C. Miller, D. A. Mantell, R. J. D. Miller,
and Y. Gao, Phys. Rev. Lett. {\bf 76}, 483 (1996).

\bibitem{Moos} G. Moos, C. Gahl, R. Fasel, M. Wolf, and T. Hertel,
Phys. Rev. Lett. {\bf 87}, 267402 (2001).

\bibitem{Gonzalez} J. Gonz\'{a}lez, F. Guinea, and M. A. H. Vozmediano,
Phys. Rev. Lett. {\bf 77}, 3589 (1996).

\bibitem{Spataru} C. D. Spataru, M. A. Cazalilla, A. Rubio, L. X. Benedict,
P. M. Echenique, and S. G. Louie, Phys. Rev. Lett. {\bf 87}, 246405 (2001).

\bibitem{Grimvall} G. Grimvall, {\it The Electron-Phonon Interaction
in Metals, Selected Topics in Solid State Physics}, edited by E. Wohlfarth (North-Holland, New York, 1981).

\bibitem{Park} C.-H. Park, F. Giustino, M. L. Cohen, and S. G. Louie,
Phys. Rev. Lett. {\bf 99}, 086804 (2007).

\bibitem{Calandra} M. Calandra and F. Mauri, Phys. Rev. B {\bf 76},
205411 (2007).

\bibitem{Lee} J. D. Lee, S. W. Han, and J. Inoue, Phys. Rev. Lett.
{\bf 100}, 216801 (2008).

\bibitem{Tuinstra} F. Tuinstra and J. L. Koenig, Journal of Chemical
Physics {\bf 53}, 1126 (1970).

\bibitem{Maultzsch} J. Maultzsch, S. Reich, C. Thomsen, H. Requardt,
and P. Ordej\'{o}n, Phys. Rev. Lett. {\bf 92}, 075501 (2004).

\bibitem{Mohr} M. Mohr, J. Maultzsch, E. Dobard\v{z}i\'{c}, S. Reich,
I. Milo\v{s}evi\'{c}, M. Damnjanovi\'{c}, A. Bosak, M. Krisch, and
C. Thomsen, Phys. Rev. B {\bf 76}, 035439 (2007).

\bibitem{Sugawara2007} K. Sugawara, T. Sato, S. Souma, T. Takahashi, and
H. Suematsu, Phys. Rev. Lett. {\bf 98}, 036801 (2007).

\bibitem{Leem} C. S. Leem, et al., Phys. Rev. Lett. {\bf 100}, 016802 (2008).

\bibitem{Bostwick} A. Bostwick, T. Ohta, T. Seyller, K. Horn, and
E. Rotenberg, Nat. Phys. {\bf 3}, 36 (2007).

\bibitem{McClure} J. W. McClure, Phys. Rev. {\bf 108}, 612 (1957).

\bibitem{Slonczewski} J. C. Slonczewski and P. R. Weiss, Phys. Rev.
{\bf 109}, 272 (1958).

\bibitem{Zunger} A. Zunger, Phys. Rev. B {\bf 17}, 626 (1978).

\bibitem{Tatar} R. C. Tatar and S. Rabii, Phys. Rev. B {\bf 25},
4126 (1982).

\bibitem{Charlier} J. C. Charlier, J. P. Michenaud, and X. Gonze,
Phys. Rev. B {\bf 46}, 4531 (1992).

\bibitem{Law1983} A. R. Law, J. J. Barry, and H. P. Hughes, Phys.
Rev. B {\bf 28}, 5332(R) (1983).

\bibitem{Takahashi} T. Takahashi, H. Tokailin, and T. Sagawa, Phys.
Rev. B {\bf 32}, 8317 (1985).

\bibitem{Law1986} A. R. Law, M. T. Johnson, and H. P. Hughes, Phys.
Rev. B {\bf 34}, 4289 (1986).

\bibitem{Collins} I. R. Collins, P. T. Andrews, and A. R. Law, Phys.
Rev. B {\bf 38}, 13348 (1988).

\bibitem{Maeda} F. Maeda, T. Takahashi, H. Ohsawa, S. Suzuki, and
H. Suematsu, Phys. Rev. B {\bf 37}, 4482 (1988).

\bibitem{Vos} M. Vos, P. Storer, S. A. Canney, A. S. Kheifets,
I. E. McCarthy, and E. Weigold, Phys. Rev. B {\bf 50}, 5635 (1994).

\bibitem{Shirley} E. L. Shirley, L. J. Terminello, A. Santoni, and
F. J. Himpsel, Phys. Rev. B {\bf 51}, 13614 (1995).

\bibitem{Heske} C. Heske, R. Treusch, F. J. Himpsel, S. Kakar,
L. J. Terminello, H. J. Weyer, and E. L. Shirley, Phys. Rev. B {\bf 59}, 4680 (1999).

\bibitem{Strocov} V. N. Strocov, A. Charrier, J. M. Themlin, M. Rohlfing,
R. Claessen, N. Barrett, J. Avila, J. Sanchez, and M. C. Asensio,
Phys. Rev. B {\bf 64}, 075105 (2001).

\bibitem{Kihlgren} T. Kihlgren, T. Balasubramanian, L. Walld\'{e}n, and
R. Yakimova, Phys.Rev. B {\bf 66}, 235422 (2002).

\bibitem{Zhou2005} S. Y. Zhou, G. H. Gweon, C. D. Spataru, J. Graf,
D. H. Lee, S. G. Louie, and A. Lanzara, Phys. Rev. B {\bf 71}, 161403(R) (2005).

\bibitem{Sugawara2006} K. Sugawara, T. Sato, S. Souma, T. Takahashi, and
H. Suematsu, Phys. Rev. B {\bf 73}, 045124 (2006).

\bibitem{Zhou2006-1} S. Y. Zhou, et al., Nat. Phys. {\bf 2}, 595 (2006).


\bibitem{Zhou2006-2} S. Y. Zhou, G. H. Gweon, and A. Lanzara,
Annals of Physics (Leipzig) {\bf 321}, 1730 (2006).

\bibitem{Binosi} D. Binosi and L. Theu{\ss}l, Comp. Phys. Comm. {\bf 161}, 76.

\bibitem{Mahan}  G. D. Mahan, {\it Many-Particle Physics}, 3rd ed.
(Kluwer Academic/Plenum Publishers, New York, 2000).

\bibitem{Feuerbacher} B. Feuerbacher and B. Fitton, Phys. Rev. Lett.
{\bf 26}, 840 (1971).

\bibitem{Balasubramanian} T. Balasubramanian, E. Jensen, X. L. Wu and
S. L. Hulbert, Phys. Rev. B {\bf 57}, 6866 (1998).

\bibitem{Ohta} T. Ohta, A. Bostwick, T. Seyller, K. Horn, and
E. Rotenberg, Sience {\bf 313}, 951 (2006).

\bibitem{Tanuma} S. Tanuma, C. J. Powell, and D. R. Penn,
Surf. Interface Anal. {\bf 17}, 911 (1991).

\end{thebibliography}
\end{document}